\documentclass[reprint,amsmath,amssymb,aps,floatfix, superscriptaddress]{revtex4-2}

\usepackage{graphicx}
\usepackage{bm}
\usepackage[english]{babel}
\usepackage[utf8]{inputenc}
\usepackage{mathtools}
\usepackage{physics}
\usepackage{xcolor}
\usepackage{graphicx}
\usepackage[left=23mm,right=13mm,top=35mm,columnsep=15pt]{geometry} 
\usepackage{adjustbox}
\usepackage{placeins}
\usepackage[T1]{fontenc}
\usepackage{lipsum}
\usepackage{csquotes}
\usepackage{float}
\usepackage{hyperref}

\usepackage{physics}
\usepackage{braket}
\usepackage{esvect}
\usepackage{amsmath} 
\usepackage{xcolor}
\usepackage{graphicx,epsfig,float,pst-all}
\usepackage{mathrsfs}
\usepackage{inputenc}
\usepackage{textcomp}
\usepackage{mathtools}
\usepackage[normalem]{ulem}
\usepackage{bm}
\usepackage{wrapfig}
\usepackage{hyperref}
\usepackage[version=4]{mhchem}
\usepackage{siunitx}
\usepackage{placeins}
\usepackage[title]{appendix}

\usepackage{dcolumn}
\usepackage[normalem]{ulem}
\usepackage{xcolor}
\usepackage{mathrsfs}
\usepackage{inputenc}
\usepackage[slantedGreek]{mathpazo}

\usepackage[color=yellow]{todonotes}

\begin{document}

\preprint{APS/123-QED}

\title{Relativistic coupled-cluster calculations for the molecular properties of AlX$^+$ (X: F, Cl, Br, I, At and Ts) ions}

\author{Ankush Thakur}
\email{ankush\_t@ph.iitr.ac.in}
\altaffiliation{Contributed equally to the work}
\affiliation{Department of Physics, Indian Institute of Technology Roorkee, Roorkee-247667, India}

\author{Renu Bala}
\email{balar180@gmail.com}
\altaffiliation{Contributed equally to the work}
\affiliation{Institute of Physics, Faculty of Physics, Astronomy and Informatics, Nicolaus Copernicus University, Grudziadzka 5, 87-100 Toru\'n, Poland}

\author{H. S. Nataraj}
\affiliation{Department of Physics, Indian Institute of Technology Roorkee, Roorkee-247667, India}

\author{V. S. Prasannaa}
\affiliation{Centre for Quantum Engineering, Research and Education, TCG Crest, Kolkata 700091, India}
\affiliation{Academy of Scientific and Innovative Research (AcSIR), Ghaziabad- 201002, India}

\begin{abstract}
In this article, the molecular permanent electric dipole moments and components of static dipole polarizabilities for the electronic ground state of singly charged aluminum monohalides are reported. The coupled-cluster method by considering single and double excitations (CCSD) together with relativistic Dyall basis sets have been used to carry out these molecular property calculations. The contribution from triple excitations are incorporated through perturbative triples (CCSD(T)). The results from a series of progressively larger basis sets are extrapolated to the complete basis set limit. Further, the role of correlation and relativistic effects, and also the effect of augmentation over the considered basis sets on the valence molecular properties are studied. Our results are compared with those available in the literature.\\
\begin{description}
\item[Keywords]{coupled-cluster, electric dipole moment, static dipole polarizability.}

\end{description}
\end{abstract}

\maketitle
\section{\label{sec:secI}Introduction}
The study of ultracold diatomic molecules has garnered significant interest in recent decades owing to their wide range of potential applications~\cite{Carr_2009,Chin_2009,Jin_2012,Softley_2023}, including fundamental physics~\cite{Mitra_2022}, quantum chemistry~\cite{Martin_2009}, quantum simulation~\cite{Cornish_2024}, etc. An important molecular property that holds significance in many of these applications is the permanent electric dipole moment (PDM) of a molecule. PDMs are important for the study of dipole-dipole interactions~\cite{Carr_2009,Tohme_2016}, quantum computing~\cite{Mille_2002,Yelin_2006,Rabl_2007}, quantum information theory~\cite{Zhang_2020}, and they are useful for exploring the characteristics of intermolecular forces~\cite{Buckingham_1959}. PDMs play a crucial role in molecular experiments aimed at detecting the electron electric dipole moment~\cite{Vutha_2010}, serving as a probe for new physics beyond the standard model of elementary particles~\cite{Fazil_2018,Sunaga_2019,Prasannaa_2019,Fleig_2017}. The importance of PDM lies in its contribution to understanding the chaining of molecules confined in a \(1\)$\mathrm{D}$  optical lattice, where the interaction strength in the process of molecular chaining is directly related to the PDM.~\cite{Wang_2006}.\\

The static electric dipole polarizability (DP) of molecules is another important property with key applications in ultracold physics. Electric DP plays a significant role in linear and non-linear optical phenomena~\cite{Buckingham_1979}, electron-molecule interactions~\cite{Lane_1980}, and also provides valuable information for studies on molecular collision processes~\cite{Boris_2021}. Both PDM and electic DP are crucial for determining the infrared and Raman molecular spectra~\cite{Banwell_1972}. Therefore, these properties are crucial in understanding the physics of ultracold molecules.\\

In the present work, we have studied the PDMs and the static electric DPs of aluminum monohalide (AlX$^+$ where $X = $  F, Cl, Br, I, At and Ts) systems. These findings could be relevant for various future applications, considering the recent focus on laser cooling of molecular ions, and theoretical studies indicating the potential for laser cooling of AlF$^+$ and AlCl$^+$ molecular ions~\cite{Kang_2017}. A few theoretical studies have been conducted on aluminum monohalide systems, in which their PDMs were reported. Klein and Rosmus~\cite{Robert_1984} have calculated the PDM for AlF$^+$ molecule using the pseudo-natural orbital configuration interaction (PNO CI) method with coupled electron pair approximation (CEPA). The study of PDM for AlF$^+$ and AlCl$^+$ molecular ions have been reported by Glenewinkel-Meyer $\textit{et al}$~\cite{Meyer_1991} using multireference configuration interaction (MRCI) method. Recently, Kang $\textit{et al}$~\cite{Kang_2017} employed the MRCI method with Davidson correction (MRCI+Q) to calculate the PDM of AlCl$^+$ molecular ion. Quite recently, the PDMs of AlX$^+$ ions have been studied by Bala $\textit{et al}$~\cite{Bala_2023} using Kramers-restricted configuration interaction method considering single and double excitations (KRCISD). However, to the best of our knowledge, there is no work in literature that reports the static electric DPs of these systems.\\

In our work, the relativistic calculations for the molecular properties have been performed using CCSD method over the Dirac-Hartree-Fock (DHF) reference state. Additionally, we have reported the molecular properties at the CCSD(T) level of theory. To carry out these calculations, we have used relativistic Dyall basis sets of double, triple and quadruple zeta quality~\cite{Dyall_2006,Dyall_2016}. Using these hierarchy of basis sets, we have extrapolated these properties to the complete basis set limit. The reported results show good agreement with the existing values of the dipole moments. We have studied the impact of electron correlation effects and relativistic effects on the molecular properties considered in this work.\\

This paper is divided into four sections. After the introduction in Section~\ref{sec:secI}, the theory and calculation details are given in Section~\ref{sec:secII}. A detailed discussion of the computed results is presented in Section~\ref{sec:secIII}. The section~\ref{sec:secIV} summarizes the current work.\\
\section{\label{sec:secII}Theory and Methodology}

\subsection{Theory}
The correlated wavefunction ($|\Psi_{CC}\rangle$) in the coupled-cluster (CC) method is expressed as the exponential of the cluster operator $\hat{T}$  acting on the DHF wavefunction ($|\Phi_0\rangle$) \cite{Shavitt_Bartlett_2009}:

\begin{equation}
    |\Psi_{CC}\rangle = e^{\hat{T}} |\Phi_0\rangle
\end{equation}

Expanding the exponential operator $e^{\hat{T}}$ into a series yields:
\begin{equation}
     |\Psi_{CC}\rangle = \Big[1+\hat{T}_1+\frac{1}{2}\hat{T}_1^{2}+\hat{T}_2+\frac{1}{3!}\hat{T}_1^{3}+\hat{T}_1\hat{T}_2+\hat{T}_3 \ldots\Big] |\Phi_0\rangle
\end{equation}

 where $\hat{T}$ (= $\hat{T}_1 + \hat{T}_2 + \hat{T}_3 \ldots$ $+ \hat{T}_n$, for an n-electron system) is the cluster operator; $\hat{T}_1, \hat{T}_2, \hat{T}_3, \ldots, \hat{T}_n$ correspond to operators that enumerate all possible single excitations, double excitations, triple excitations, \ldots, n-tuple excitations that arise from the reference DHF state. In second quantized notation, these operators can be expressed as,

\begin{equation}
    \hat{T}_{1} = \sum_{a,p} t_a^{p} \hat{a}_p^\dagger \hat{a}_a,  
\end{equation} 
\begin{equation}
    \hat{T}_{2} = \frac{1}{4} \sum_{ab,pq} t_{ab}^{pq} \hat{a}_p^\dagger \hat{a}_q^\dagger \hat{a}_b \hat{a}_a,
\end{equation} 
and so on, where the subscripts $a$, $b$, $c$, $\ldots$ represent filled spin-orbitals, and $p$, $q$, $r$, $\ldots$ denote the virtual spin-orbitals. Thus, the operator $\hat{T}_1$ denotes the annihilation of an electron from the filled spin-orbital $a$, accompanied by the creation of a virtual electron in spin-orbital $p$, and this single excitation is weighted by the amplitude $t_a^{p}$. Similarly, $\hat{T}_2$ signifies the simultaneous annihilation of a pair of electrons from the filled spin-orbitals ($a$, $b$) and their creation in the virtual spin-orbitals ($p$, $q$), with an associated amplitude, $t_{ab}^{pq}$.\\

We employ the finite-field method to compute molecular properties.In the presence of a uniform external electric field of strength $\varepsilon$, the total energy E($\varepsilon$) of a molecule can be expressed as a Taylor series expansion~\cite{Mitra_2020}:
\begin{eqnarray}
    E(\varepsilon) = E_0 - \mu \varepsilon - \frac{1}{2}\alpha \varepsilon^2 + \cdots , 
\end{eqnarray}
where
\begin{eqnarray}
    \mu = -\frac{dE(\varepsilon)}{d\varepsilon};  \hspace{0.5cm}  \alpha = -\frac{d^2E(\varepsilon)}{d\varepsilon^2}.
\end{eqnarray}
where $\mu$ is the PDM and $\alpha$ is the static DP.

\subsection{Details of calculation}

The PDM and the static electric DP ($\alpha_\parallel$ and $\alpha_\perp$) calculations reported in this work are computed at different levels of theory using DIRAC23~\cite{DIRAC} software package. We utilize a \(4\)-component wavefunction, which is expanded using distinct basis sets for both large and small components. The kinetic balance condition is applied to smaller components to prevent the wavefunction from experiencing a variational collapse into the negative energy continuum~\cite{Kenneth_2007}. The DHF Coulomb Hamiltonian is employed with the approximation proposed by Visscher \cite{Visscher1997}, wherein the contribution from the (SS$\vert$SS) integrals is replaced by an inter-atomic correction. We have used the uncontracted Dyall basis sets for the calculations. The details of the basis functions are shown in Table~\ref{table-I}. The values of equilibrium bond lengths used in this work are \cite{Bala_2023}: \(1.623\) $\si{\angstrom}$ for \ce{AlF^+}, \(2.058\) $\si{\angstrom}$ for \ce{AlCl^+}, \(2.223\) $\si{\angstrom}$ for \ce{AlBr^+}, \(2.512\) $\si{\angstrom}$ for \ce{AlI^+}, \(2.768\) $\si{\angstrom}$ for \ce{AlAt^+}, and \(2.928\) $\si{\angstrom}$ for \ce{AlTs^+}. The strength of the electric field taken as perturbation is chosen in the range of \(-1\) $\times$ $10^{-4}$ to \(1\) $\times$ $10^{-4}$ au. The aluminum atom is chosen to be at the origin of the coordinate axes. \\

\begin{table}[ht]
    \centering
    \caption{\label{table-I}Details of the basis functions.}
    \begin{tabular}{c c c}
    \hline\hline
       Atom & Basis & Basis functions \\
       \hline
       Al & dyall.v2z & 12s, 8p, 1d \\
          & dyall.v3z & 18s, 11p, 2d, 1f\\
          & dyall.v4z & 24s, 14p, 3d, 2f, 1g \\
          & s-aug-dyall.v4z & 25s, 15p, 4d, 3f, 2g \\
          \hline
       F & dyall.v2z & 10s, 6p, 1d \\
         & dyall.v3z & 14s, 8p, 2d, 1f\\
         & dyall.v4z & 18s, 10p, 3d, 2f, 1g \\
         & s-aug-dyall.v4z & 19s, 11p, 4d, 3f, 2g \\
         \hline
       Cl & dyall.v2z & 12s, 8p,1d \\
          & dyall.v3z & 18s, 11p, 2d, 1f\\
          & dyall.v4z & 24s, 14p, 3d, 2f, 1g \\
          & s-aug-dyall.v4z & 25s, 15p, 4d, 3f, 2g \\
          \hline
       Br & dyall.v2z & 15s, 11p, 7d \\
          & dyall.v3z & 23s, 16p, 10d, 1f\\
          & dyall.v4z & 30s, 21p, 13d, 2f, 1g \\
          & s-aug-dyall.v4z & 31s, 22p, 14d, 3f, 2g \\
          \hline
       I & dyall.v2z & 21s, 15p, 11d \\
         & dyall.v3z & 28s, 21p, 15d, 1f\\
         & dyall.v4z & 33s, 27p, 18d, 2f, 1g \\
         & s-aug-dyall.v4z & 34s, 28p, 19d, 3f, 2g \\
         \hline
       At & dyall.v2z & 24s, 20p, 14d, 8f \\
          & dyall.v3z & 30s, 26p, 17d, 11f \\
          & dyall.v4z & 34s, 31p, 21d, 14f, 1g \\
          & s-aug-dyall.v4z & 35s, 32p, 22d, 15f, 2g \\
          \hline
       Ts & dyall.v2z & 26s, 23p, 17d, 10f \\
          & dyall.v3z & 30s, 29p, 20d, 13f\\
          & dyall.v4z & 35s, 35p, 24d, 16f, 1g \\
          & s-aug-dyall.v4z & 36s, 36p, 25d, 17f, 2g \\
       \hline\hline
    \end{tabular}
    \label{tab:my_label}
\end{table}
By considering the z-axis as the internuclear axis of the molecule, we obtain two components of DP: one along ($\alpha_\parallel$ $\equiv$ $\alpha_{zz}$) and the other perpendicular ($\alpha_\perp$ $\equiv$ $\alpha_{xx}$ $\equiv$ $\alpha_{yy}$) to the direction of internuclear axis. The average and the anisotropic polarizabilities, ($\bar{\alpha}$) and ($\gamma$) respectively, are given by, 

\begin{equation}
    \bar{\alpha}=(\alpha_\parallel + 2\alpha_\perp)/3
\end{equation}

\begin{equation}
    \gamma=\alpha_\parallel - \alpha_\perp.
\end{equation}

The cutoff energy of \(12\)$E_h$ is consistently set for all molecules to limit the higher virtual orbitals, thereby reducing computational expenses. The details of the active electrons and virtual orbitals for different basis sets are shown in Table~\ref{table-II}. 

\begin{table}[ht]
    \centering
    \caption{\label{table-II}Details of the number of active electrons and virtual orbitals for \ce{AlX^+} (X: F, Cl, Br, I, At and Ts) molecular systems in different basis sets.}
    \begin{tabular}{c c c c}
    \hline\hline
    Molecule & Active electrons & Basis & Virtual orbitals \\
    \hline
    \ce{AlF^+} & 15 & dyall.v2z & 35 \\
               &    & dyall.v3z & 63 \\
               &    & dyall.v4z & 102 \\
               &    & s-aug-dyall.v4z & 143 \\
    \hline           
    \ce{AlCl^+} & 15 & dyall.v2z & 32 \\
                &    & dyall.v3z & 64 \\
                &    & dyall.v4z & 115 \\
                &    & s-aug-dyall.v4z & 165 \\
    \hline
    \ce{AlBr^+} & 25 & dyall.v2z & 37 \\
                &    & dyall.v3z & 69 \\
                &    & dyall.v4z & 120 \\
                &    & s-aug-dyall.v4z & 170 \\
    \hline
    \ce{AlI^+} & 25 & dyall.v2z & 42 \\
               &    & dyall.v3z & 74 \\
               &    & dyall.v4z & 125 \\
               &    & s-aug-dyall.v4z & 170 \\
    \hline
    \ce{AlAt^+} & 29 & dyall.v2z & 49 \\
                &    & dyall.v3z & 85 \\
                &    & dyall.v4z & 138 \\
                &    & s-aug-dyall.v4z & 188 \\
    \hline
    \ce{AlTs^+} & 29 & dyall.v2z & 56 \\
                &    & dyall.v3z & 84 \\
                &    & dyall.v4z & 137 \\
                &    & s-aug-dyall.v4z & 187 \\
    \hline\hline
    \end{tabular}
    \label{tab:my_label}
\end{table}

The results calculated systematically using three progressively larger basis sets, are extrapolated to the complete basis set CBS limit using a function of the form~\cite{Peterson_1994}:
\begin{equation} \label{eq:9}
    f(x) = f_{CBS} + B e^{-(x-1)} + C e^{-(x-1)^2}
\end{equation}
where $B$ and $C$ are constant parameters, \(x\) = $2$, $3$, $4$ is the cardinal number for double, triple and quadruple zeta basis sets respectively, $f(x)$ represents the value of the property calculated with the basis set characterized by the cardinal number \(x\), and $f_{CBS}$ denotes the value of the property of interest at the CBS limit. This approach for computing the CBS limit for molecular properties has been applied in the literature~\cite{Feller_2000,Bala_2018}.\\

\section{\label{sec:secIII}Results and Discussion}
The computed results for the molecular properties are discussed in the upcoming subsections. We also investigate the relative percentage difference in the values of PDM and DP for the AlX$^+$ molecular systems due to electron correlation and relativistic effects. For the relative percentage change due to correlation effects, we define $\delta^{\text{corr}}_{\text{P}}$ as:
\begin{equation}
     \delta^{\text{corr}}_{\text{P}} = \left( \frac{{P_\text{CCSD(T)} - P_\text{DHF}}}{{P_\text{DHF}}} \times 100 \right)\%
\end{equation}
This indicates the relative percentage of correlation effects contributing to the property $P$.\\
For the relative percentage change due to relativistic effects, we define $\delta^{\text{rel}}_{\text{P}}$ as:
\begin{equation}
     \delta^{\text{rel}}_{\text{P}} = \left( \frac{{P_\text{Rel} - P_\text{Non-Rel}}}{{P_\text{Non-Rel}}} \times 100 \right)\% 
\end{equation}

\begin{table}[htbp]
    \caption{\label{table-III}Magnitude of permanent electric dipole moments (in Debye) for \ce{AlX^+} (X: F, Cl, Br, I, At and Ts) molecular systems at different levels of correlation and basis sets.}  
    \begin{ruledtabular}
    \begin{tabular}{c c c c c}
       Molecule & Basis & Method &$\mu$ & Ref. \\
       \hline
       \ce{AlF^+} & dyall.v2z & DHF & 2.821 & This work \\
                  &           & CCSD & 2.436 & This work \\
                  &           & CCSD(T) & 2.327 & This work\\
                  \\
                  & dyall.v3z & DHF & 2.569 & This work \\
                  &           & CCSD & 2.296 & This work \\
                  &           & CCSD(T) & 2.210 & This work \\
                  \\
                  & dyall.v4z & DHF & 2.550 & This work \\
                  &           & CCSD & 2.300 & This work\\
                  &           & CCSD(T) & 2.220 & This work \\
                  \\
                  & \textbf{CBS} & DHF & 2.546 &  This work\\
                                && CCSD & 2.308 & This work\\
                                && \textbf{CCSD(T)} & \textbf{2.231} & This work\\
                                \\
                  &           & DHF & 2.56 &\cite{Bala_2023}\\
                  &          & KRCISD  & 2.33 &\cite{Bala_2023}\\
                  &           & MRCI    & 2.40 & \cite{Meyer_1991}\\
                  &           & HF & 2.39 &\cite{Robert_1984}\\
                  &           & PNO-CEPA & 2.29 & \cite{Robert_1984}\\
        \hline\hline
        \ce{AlCl^+} & dyall.v2z & DHF & 0.937 & This work\\
                  &           & CCSD & 0.279 & This work\\
                  &           & CCSD(T) & 0.121 & This work\\
                  \\
                  & dyall.v3z & DHF & 0.768 & This work \\
                  &           & CCSD & 0.293 & This work \\
                  &           & CCSD(T) & 0.155 & This work \\
                  \\
                  & dyall.v4z & DHF & 0.746 & This work \\
                  &           & CCSD & 0.346 & This work \\
                  &           & CCSD(T) & 0.217 & This work \\
                  \\
                  & \textbf{CBS} & DHF & 0.737 & This work\\
                                && CCSD & 0.381 & This work\\
                                && \textbf{CCSD(T)} & \textbf{0.258} & This work\\
                                \\
                  & & DHF & 0.76 &\cite{Bala_2023}\\
                   &          & KRCISD  & 0.47 &\cite{Bala_2023}\\
                  &           & MRCI    & 0.27 &\cite{Kang_2017}\\
                  &           & MRCI    & 0.19 &\cite{Meyer_1991}\\
        \hline\hline
        \ce{AlBr^+} & dyall.v2z & DHF & 0.663 & This work \\
                  &           & CCSD & 0.129 & This work \\
                  &           & CCSD(T) & 0.307 & This work \\
                  \\
                  & dyall.v3z & DHF & 0.493 & This work \\
                  &           & CCSD & 0.126 & This work \\
                  &           & CCSD(T) & 0.285 & This work \\
                  \\
                  & dyall.v4z & DHF & 0.486 & This work \\
                  &           & CCSD & 0.038 & This work \\
                  &           & CCSD(T) & 0.192 & This work \\
                  \\
                  & \textbf{CBS} & DHF & 0.487 & This work\\
                                && CCSD & 0.021 & This work\\
                                && \textbf{CCSD(T)} & \textbf{0.130} & This work \\

            \end{tabular}
    \label{tab:my_label}
    \end{ruledtabular}
\end{table}
\setcounter{table}{2}

\begin{table}[htbp]
    \caption{ \label{table-III} Continued...}
    \begin{ruledtabular}
    \begin{tabular}{c c c c c}
                  &           & DHF  & 0.50 &\cite{Bala_2023}\\
                  &           & KRCISD & 0.07 &\cite{Bala_2023}\\
                  \hline \hline
     \ce{AlI^+} & dyall.v2z & DHF & 1.269 & This work\\
                  &           & CCSD & 2.434 & This work \\
                  &           & CCSD(T) & 2.636 & This work \\
                  \\
                  & dyall.v3z & DHF & 1.493 & This work \\
                  &           & CCSD & 2.472 & This work \\
                  &           & CCSD(T) & 2.654 & This work \\
                  \\
                  & dyall.v4z & DHF & 1.501 & This work \\
                  &           & CCSD & 2.342 & This work \\
                  &           & CCSD(T) & 2.524 & This work \\
                  \\
                   & \textbf{CBS} & DHF & 1.499 & This work\\
                                && CCSD & 2.253 & This work\\
                                && \textbf{CCSD(T)} & \textbf{2.435} & This work \\
                                \\
                  &&DHF & 1.49 &\cite{Bala_2023}\\
                  &           & KRCISD  & 2.16 &\cite{Bala_2023}\\
            \hline\hline
     \ce{AlAt^+} & dyall.v2z & DHF & 4.033 & This work \\
                  &           & CCSD & 4.281 & This work\\
                  &           & CCSD(T) & 4.357 & This work \\
                  \\
                  & dyall.v3z & DHF & 4.292 & This work\\
                  &           & CCSD & 4.396 & This work \\
                  &           & CCSD(T) & 4.428 & This work\\
                  \\
                  & dyall.v4z & DHF & 4.304 & This work \\
                  &           & CCSD & 4.299 & This work\\
                  &           & CCSD(T) & 4.316 & This work\\
                  \\
                   & \textbf{CBS} & DHF & 4.303 & This work\\
                                && CCSD & 4.229 & This work\\
                                && \textbf{CCSD(T)} & \textbf{4.238} & This work\\
                                \\
                  &&DHF&4.31 &\cite{Bala_2023}\\
                  &           & KRCISD  &4.24 &\cite{Bala_2023}\\
        \hline\hline
        \ce{AlTs^+} & dyall.v2z & DHF & 6.444 & This work \\
                  &           & CCSD & 6.623 & This work\\
                  &           & CCSD(T) & 6.802 & This work\\
                  \\
                  & dyall.v3z & DHF & 6.696 & This work \\
                  &           & CCSD & 6.687 & This work\\
                  &           & CCSD(T) & 6.826 & This work\\
                  \\
                  & dyall.v4z & DHF & 6.722 & This work \\
                  &           & CCSD & 6.581 & This work \\
                  &           & CCSD(T) & 6.697 & This work \\
                  \\
                   & \textbf{CBS} & DHF & 6.731 & This work \\
                                && CCSD & 6.507 & This work\\
                                && \textbf{CCSD(T)} & \textbf{6.609} & This work \\
                                \\
                  &&DHF& 6.76 &\cite{Bala_2023}\\
                  &           & KRCISD  & 6.91 &\cite{Bala_2023}\\
    \end{tabular}
 \end{ruledtabular}  
\end{table}

All the reported values for the molecular properties have been rounded off to three decimal places in this work. Our final results at the CBS limit are shown in bold font in Table~\ref{table-III} and Table~\ref{table-IV}.

\subsection{Permanent electric dipole moments}
\begin{figure}[hbp]
    \includegraphics[width=9.3cm,height=6.5cm]{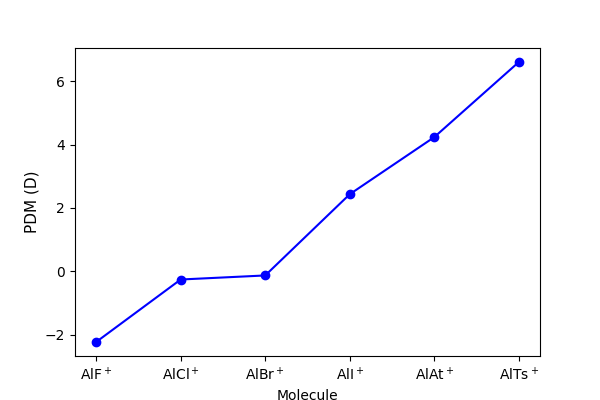}
    \caption{\label{fig:FIG1}The CBS values for molecular PDMs for the ground state of \ce{AlX^+} (X: F, Cl, Br, I, At and Ts) molecular systems at CCSD(T) level of theory. }
\end{figure}

Table~\ref{table-III} shows the results for the magnitude of the PDMs of singly charged aluminum monohalides calculated in the present work, compared with the results available in the literature. The PDM is negative for AlF$^+$ and AlCl$^+$, whereas it is positive for the remaining molecules, and it increases as one progresses from lighter to heavier systems, as shown in Figure~\ref{fig:FIG1}. As molecules become heavier, the PDM tends to increase due to their larger atomic size, resulting in stronger charge separation. \\

For all the molecular ions, the PDM values at the DHF level show good agreement with those available in the literature~\cite{Bala_2023,Robert_1984}. For the AlF$^+$ molecule, the dipole moment value at the CCSD(T) level compares well, differing by approximately \(4.9\)\% from that at the KRCISD level, as reported in Ref.~\cite{Bala_2023}, all under a similar standard of basis set and equilibrium bond length but with different active spaces.\\

Glenewinkel-Meyer $\textit{et al}$~\cite{Meyer_1991} calculated the PDM of the AlF$^+$ molecule using complete active space self-consistent-field  (CASSCF) and internally contracted MRCI methods. They employed Gaussian-type orbital basis sets, including Partridge's \(17\)$s$, \(12\)$p$ for Al and Dunning's quadruple zeta for F. Subsequent MRCI calculations, included in their work, comprised all single and double excitations relative to the CASSCF reference functions, incorporating correlations for all valence electrons. They obtain a PDM of \(2.40\) Debye ($\mathrm{D}$), indicating a \(0.17\) $\mathrm{D}$ increase compared to our result of $2.231$ ($\mathrm{D}$) (at CCSD(T) level). In addition to the difference in the methods, their basis set size and selection of active space are considerably smaller in comparison to ours.\\

Klein and Rosmus \cite{Robert_1984} have reported the PDM for AlF$^+$ molecule. They employed PNO CI method with CEPA approximation, utilizing a GTO basis set with \(12\)$s$, \(9\)$p$, and \(2\)$d$ functions for Al, and \(10\)$s$, \(6\)$p$, and \(2\)$d$ functions for F. Within the CI level, they considered single and double excitations relative to the reference HF state. To assess the impact of excitations higher than single and double levels, they employed the CEPA approximation in their work. In our work, the value of PDM at the CCSD(T) level exceeds that reported in their study at the PNO-CEPA level by \(2.6\)\%. This difference could be due to the different methods and large basis set employed in our work.\\

For the AlCl$^+$ molecule, the value of PDM reported in our work using the CCSD(T) method is \(0.012\) $\mathrm{D}$ smaller and \(0.068\) $\mathrm{D}$ larger than those reported in Ref.~\cite{Kang_2017} and Ref.~\cite{Meyer_1991}, respectively, at MRCI level of theory. Our value of PDM at the CCSD(T) level is smaller by \(0.212\) $\mathrm{D}$ compared to that reported in Ref.~\cite{Bala_2023} at the KRCISD level.\\
\begin{figure}[h]
    \includegraphics[width=9.3cm,height=6.5cm]{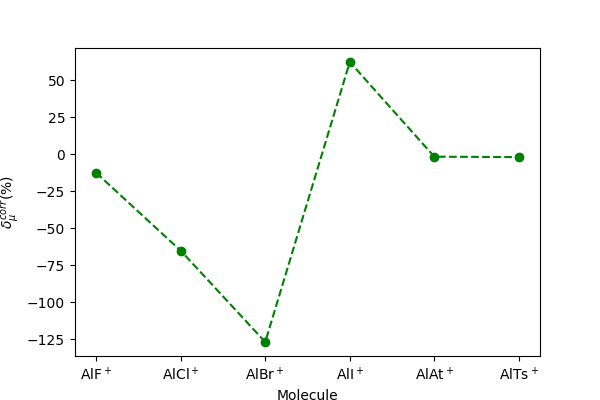}
    \caption{\label{fig:FIG2}Figure showing the relative percentage changes in the values of PDMs in the \ce{AlX^+} (X: F, Cl, Br, I, At and Ts) molecular systems due to the electron correlation effects. }
\end{figure}

\begin{figure}[]
    \includegraphics[width=9.3cm,height=6.5cm]{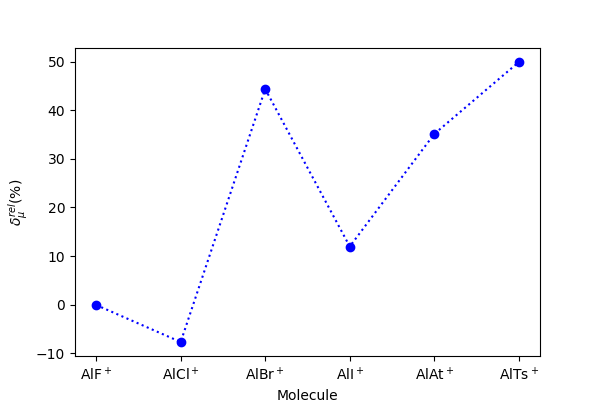}
    \caption{\label{fig:FIG3}Figure showing the relative percentage changes in the values of PDMs in the \ce{AlX^+} (X: F, Cl, Br, I, At and Ts) molecular systems due to the relativistic effects. }
\end{figure}

Bala $\textit{et al}$ \cite{Bala_2023} had computed the PDM of AlBr$^+$, AlI$^+$, AlAt$^+$, and AlTs$^+$ molecules at the DHF and KRCISD levels of theory using dyall.v4z basis sets. Our results show good agreement with those reported in their work. Our PDM values at CCSD(T) levels show increments of \(0.06\) $\mathrm{D}$, \(0.275\) $\mathrm{D}$, \(0.002\) $\mathrm{D}$, and \(0.301\) $\mathrm{D}$ for AlBr$^+$, AlI$^+$, AlAt$^+$, and AlTs$^+$, respectively, in comparison to those reported by them at KRCISD level. These differences in the results could be due to the ability of CC to capture more electron correlation effects than CI~\cite{Bartlett_2007,Sahoo_2017}.\\

Electron correlation significantly affects the accuracy of final values, with its impact reaching maximum in the CCSD(T) calculations for AlI$^+$, where the correlation fraction (denoted as $\delta^{\text{corr}}_{\text{$\mu$}}$ in Figure~\ref{fig:FIG2}) constitutes approximately \(62\)\%. For AlF$^+$ and AlCl$^+$ molecular ions, the values of PDM decrease from the DHF method to the CCSD(T) method, whereas AlI$^+$ exhibits a contrasting trend with an increase in value. The PDM value of the AlBr$^+$, AlAt$^+$, and AlTs$^+$ molecular ions decreases from the DHF to the CCSD method, followed by an increase in the CCSD(T) method. In summary, electron correlation effects leads to an increase in the PDM value for AlI$^+$ molecular ion, whereas it results in a decrease in the PDM value for the remaining molecular ions. Figure~\ref{fig:FIG2} illustrates these observed trends. \\ 

We now compare the PDM results obtained from both non-relativistic and relativistic methods, using the data presented in Table~\ref{table-V}. The HF Coulomb Hamiltonian has been employed for non-relativistic calculations. In Figure~\ref{fig:FIG3}, we have depicted the fractional percentage difference in PDM values due to relativistic effects (denoted as $\delta^{\text{rel}}_{\text{$\mu$}}$). The effect of relativity seems to decrease the value of PDM for lighter molecular ions, such as AlF$^+$ and AlCl$^+$, and to increase the value for other molecular ions. We observe that the significance of relativistic effects is most notable in the heavier molecules, naturally as anticipated, with the PDM of AlAt$^+$ and AlTs$^+$ molecular ions showing the change of \(35\)\% and \(50\)\%, respectively, with the inclusion of relativistic effects. \\

As the molecular properties studied in this work are valence properties, the addition of diffuse functions to the basis sets can play an important role in the accuracy of results~\cite{Berredo_2011}. Therefore, we have also studied the effect of augmentation of basis sets on the results of PDM by considering singly augmented quadruple zeta basis sets, as shown in Table~\ref{table-VI}. The difference in our final results of PDMs between the un-augmented and augmented basis sets is about \(0.001\) $\mathrm{D}$ for AlF$^+$, \(0.005\) $\mathrm{D}$ for AlCl$^+$, \(0.007\) $\mathrm{D}$ for AlBr$^+$, \(0.012\) $\mathrm{D}$ for AlI$^+$, \(0.007\) $\mathrm{D}$ for AlAt$^+$, and \(0.01\) $\mathrm{D}$ for AlTs$^+$ at the CCSD(T) level. Thus, the augmentation of the basis sets does not alter the results of the PDMs by more than \(3.6\)\% . \\

In order to understand the errors in our calculation, we have considered the heavy AlAt$^+$ molecule as our representative system. One potential source of uncertainty in our calculation may stem from the exclusion of more number of active electrons. To address this point, we have incorporated \(49\) active electrons rather than \(29\) in our active space. This results in a \(1.7\)\% change in the value of PDM. We have found a \(1.9\)\% change in the PDM due to the increase in the virtual cut-off from \(12\) $E_h$ to \(25\) $E_h$ and a \(1.7\)\% change with the addition of augmented functions to the basis sets. We now consider the error resulting from the missing triple excitations in the CCSD(T) method. This method is widely recognized as the gold standard for molecular property calculations~\cite{Maitra_2017}, and we anticipate that the error due to missing triple excitations would be significantly lower than the percentage difference between the CCSD and CCSD(T) values, which we found to be \(0.2\)\%. All these variations result in a total change of about \(5.5\)\% in the PDM value. 

\subsection{Static electric dipole polarizabilities}
\begin{table*}[t]
    \caption{\label{table-IV}Components of static dipole polarizability (in au) for \ce{AlX^+} (X: F, Cl, Br, I, At and Ts) molecular systems at different levels of correlation, computed in this work.}
    \begin{ruledtabular}
    \begin{tabular}{c c c c c c c}
       Molecule & Basis & Method & $\alpha_\parallel$ & $\alpha_\perp$ & $\bar{\alpha}$ & $\gamma$ \\
       \hline
       \ce{AlF^+} & dyall.v2z & DHF & 16.256 & 20.240 & 18.912 & -3.984 \\
                  &           & CCSD & 18.889 & 20.427 & 19.914 & -1.538 \\
                  &           & CCSD(T) & 19.680 & 20.512 & 20.235 & -0.832 \\
                  \\
                  & dyall.v3z & DHF & 16.723 & 21.300 & 19.774 & -4.577 \\
                  &           & CCSD & 18.981 & 21.393 & 20.589 & -2.412 \\
                  &           & CCSD(T) & 19.654 & 21.485 & 20.875 & -1.831 \\
                  \\
                  & dyall.v4z & DHF & 16.857 & 21.678 & 20.071 & -4.821 \\
                  &           & CCSD & 19.080 & 21.637 & 20.785 & -2.557 \\
                  &           & CCSD(T) & 27.610 & 17.180 & 20.657 & 10.430 \\
                  \\
                  &\textbf{CBS} & DHF & - & - & 20.242 & -4.965 \\
                               && CCSD & - & - & 20.894 & -2.625 \\
                                && \textbf{CCSD(T)} & - & - & \textbf{20.487} & \textbf{18.763} \\
        \hline\hline
        \ce{AlCl^+} & dyall.v2z & DHF & 36.275 & 26.144 & 29.521 & 10.131 \\
                  &           & CCSD & 41.563 & 26.572 & 31.569 & 14.991 \\
                  &           & CCSD(T) & 42.800 & 26.689 & 32.059 & 16.111 \\
                  \\
                  & dyall.v3z & DHF & 38.234 & 29.406 & 32.349 & 8.828 \\
                  &           & CCSD & 43.007 & 29.636 & 34.093 & 13.371 \\
                  &           & CCSD(T) & 44.254 & 29.756 & 34.589 & 14.498 \\
                  \\
                  & dyall.v4z & DHF & 38.549 & 30.480 & 33.170 & 8.069 \\
                  &           & CCSD & 42.996 & 30.551 & 34.699 & 12.445 \\
                  &           & CCSD(T) & 43.706 & 29.996 & 34.566 & 13.710 \\
                  \\
                   &\textbf{CBS} & DHF & - & - & 33.627 & 7.601\\
                                && CCSD & - & - & 35.021 & 11.875 \\
                                && \textbf{CCSD(T)} & - & - & \textbf{34.462} & \textbf{13.233}\\
        \hline\hline
        \ce{AlBr^+} & dyall.v2z & DHF & 48.506 & 29.701 & 35.969 & 18.805 \\
                  &           & CCSD & 54.899 & 30.235 & 38.456 & 24.664 \\
                  &           & CCSD(T) & 56.194 & 30.361 & 38.972 & 25.833 \\
                  \\
                  & dyall.v3z & DHF & 51.314 & 34.368 & 40.017 & 16.946 \\
                  &           & CCSD & 57.271 & 34.662 & 42.198 & 22.609 \\
                  &           & CCSD(T) & 58.569 & 34.806 & 42.727 & 23.763 \\
                  \\
                  & dyall.v4z & DHF & 51.742 & 35.732 & 41.069 & 16.010 \\
                  &           & CCSD & 57.289 & 35.820 & 42.976 & 21.469 \\
                  &           & CCSD(T) & 58.411 & 35.788 & 43.329 & 22.623 \\
                  \\
                   &\textbf{CBS} & DHF & - & - & 41.640 & 15.441 \\
                                && CCSD & - & - & 43.372 & 20.769 \\
                                && \textbf{CCSD(T)} & - & - & \textbf{43.606} & \textbf{21.924}\\
            \end{tabular}
    \label{tab:my_label}
    \end{ruledtabular}
\end{table*}
\setcounter{table}{3}
\begin{table*}[t]
    \caption{ \label{table-IV} Continued...}
    \begin{ruledtabular}
    \begin{tabular}{c c c c c c c}
      \ce{AlI^+} & dyall.v2z & DHF & 77.196 & 37.036 & 50.423 & 40.160 \\
                  &           & CCSD & 86.241 & 38.072 & 54.128 & 48.169 \\
                  &           & CCSD(T) & 87.056 & 38.173 & 54.467 & 48.883 \\
                  \\
                  & dyall.v3z & DHF & 79.804 & 44.019 & 55.947 & 35.785 \\
                  &           & CCSD & 88.256 & 44.432 & 59.040 & 43.824 \\
                  &           & CCSD(T) & 88.978 & 44.546 & 59.357 & 44.432 \\
                  \\
                  & dyall.v4z & DHF & 80.426 & 45.804 & 57.345 & 34.622 \\
                  &           & CCSD & 88.449 & 45.835 & 60.040 & 42.614 \\
                  &           & CCSD(T) & 89.195 & 46.023 & 60.414 & 43.172 \\
                  \\
                   &\textbf{CBS} & DHF & - & - & 58.099 & 33.988 \\
                                && CCSD & - & - & 60.546 & 41.947 \\
                                && \textbf{CCSD(T)} & - & - & \textbf{60.959} & \textbf{42.474} \\
                    \hline\hline
     \ce{AlAt^+} & dyall.v2z & DHF & 103.810 & 57.430 & 72.890 & 46.380 \\
                  &           & CCSD & 100.210 & 58.220 & 72.217 & 41.990 \\
                  &           & CCSD(T) & 102.700 & 50.580 & 67.953 & 52.120 \\
                  \\
                  & dyall.v3z & DHF & 107.860 & 59.279 & 75.473 & 48.581 \\
                  &           & CCSD & 102.620 & 57.900 & 72.807 & 44.720 \\
                  &           & CCSD(T) & 104.960 & 53.210 & 70.460 & 51.750 \\
                  \\
                  & dyall.v4z & DHF & 108.780 & 60.480 & 76.580 & 48.300 \\
                  &           & CCSD & 103.230 & 58.690 & 73.537 & 44.540 \\
                  &           & CCSD(T) & 105.880 & 54.620 & 71.707 & 51.260 \\
                  \\
                   &\textbf{CBS} & DHF & - & - & 77.239 & 48.033 \\
                                && CCSD & - & - & 74.010 & 44.323 \\
                                && \textbf{CCSD(T)} & - & - & \textbf{72.464} & \textbf{50.941} \\
        \hline\hline
        \ce{AlTs^+} & dyall.v2z & DHF & 133.630 & 87.051 & 102.577 & 46.579 \\
                  &           & CCSD & 132.190 & 60.520 & 84.410 & 71.670 \\
                  &           & CCSD(T) & 133.859 & 3.810 & 47.160 & 130.049 \\
                  \\
                  & dyall.v3z & DHF & 134.811 & 89.620 & 104.684 & 45.191 \\
                  &           & CCSD & 132.061 & 58.529 & 83.040 & 73.532 \\
                  &           & CCSD(T) & 133.520 & 31.930 & 65.793 & 101.590 \\
                  \\
                  & dyall.v4z & DHF & 135.630 & 90.080 & 105.263 & 45.550 \\
                  &           & CCSD & 131.010 & 58.663 & 82.779 & 72.347 \\
                  &           & CCSD(T) & 132.180 & 54.477 & 80.378 & 77.703 \\
                  \\
                   &\textbf{CBS} & DHF & - & - & 105.581 & 45.841 \\
                                && CCSD & - & - & 82.650 & 71.480 \\
                                && \textbf{CCSD(T)} & - & - & \textbf{89.599} & \textbf{62.530} \\
    \end{tabular}
 \end{ruledtabular}   
\end{table*}

\begin{table*}[ht]
    \caption{\label{table-V}Comparison of the magnitudes of PDMs (in Debye) and components of static dipole polarizability (in au) obtained from relativistic and non-relativistic calculations at different levels of correlation using dyall.v4z basis sets.}
    \begin{ruledtabular}
    \begin{tabular}{c c c c c c c c}
      &  & \multicolumn{3}{c}{Non-relativistic} & \multicolumn{3}{c}{Relativistic} \\ \cline{3-5} 
      \cline{6-8}
       Molecule & Method & $\mu$ & $\bar{\alpha}$ & $\gamma$  & $\mu$ & $\bar{\alpha}$ & $\gamma$ \\
       \hline
       \ce{AlF^+} & HF/DHF & 2.547 & 20.153 & -4.957 & 2.550 & 20.071 & -4.821 \\
                  & CCSD & 2.301 & 20.851 & -2.886 & 2.300 & 20.785 & -2.557 \\
                  & CCSD(T) & 2.221 & 10.723 & 13.275 & 2.220 & 20.657 & 10.430  \\
        \hline
        \ce{AlCl^+} & HF/DHF & 0.757 & 33.187 & 7.828 & 0.746 & 33.170 & 8.069 \\
                    & CCSD & 0.363 & 35.053 & 12.270 & 0.346 & 34.699 & 12.445 \\
                    & CCSD(T) & 0.235 & 34.201 & 15.372 & 0.217 & 34.566 & 13.710 \\
        \hline
        \ce{AlBr^+} & HF/DHF & 0.531 & 40.994 & 15.457 & 0.486 & 41.069 & 16.010 \\
                    & CCSD & 0.022 & 43.055 & 21.965 & 0.038 & 42.976 & 21.469 \\
                    & CCSD(T) & 0.133 & 43.489 & 23.337 & 0.192 & 43.329 & 22.623 \\
        \hline
        \ce{AlI^+} & HF/DHF & 1.295 & 56.356 & 31.709 & 1.501 & 57.345 & 34.622 \\
                   &  CCSD & 2.056 & 58.247 & 36.354 & 2.342 & 60.040 & 42.614 \\
                   & CCSD(T) & 2.254 & 58.785 & 37.402 & 2.524 & 60.414 & 43.172 \\
        \hline
     \ce{AlAt^+}  & HF/DHF & 2.053 & 70.403 & 52.420 & 4.304 & 76.580 & 48.300 \\
                  & CCSD & 3.036 & 72.870 & 57.960 & 4.299 & 73.537 & 44.540 \\
                  & CCSD(T) & 3.194 & 72.623 & 56.140 & 4.316 & 71.707 & 51.260 \\
        \hline
    \ce{AlTs^+} &  HF/DHF & 3.380 & 85.077 & 67.970 & 6.722 & 105.263 & 45.550 \\
                &  CCSD & 4.326 & 81.350 & 56.669 & 6.581 & 82.779 & 72.347 \\
                &  CCSD(T) & 4.467 & 77.017 & 60.829 & 6.697 & 80.378 & 77.703 \\
    \end{tabular}
    \label{tab:my_label}
 \end{ruledtabular}   
\end{table*}

\begin{figure}[hbp]
    \includegraphics[width=9.3cm,height=6.5cm]{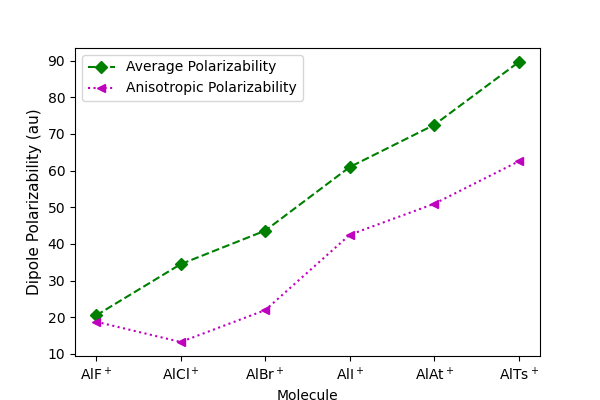}
    \caption{\label{fig: FIG4}The CBS values for the average and anisotropic component of polarizability for the ground state of \ce{AlX^+} (X: F, Cl, Br, I, At and Ts) molecular systems at CCSD(T) level of theory. }
\end{figure}
The computed results for the components of static DP are collected in Table~\ref{table-IV}. The results of electric DPs for these molecular ions are not available in the literature, to the best of our knowledge, for comparison. We prioritize analyzing the trends of $\bar{\alpha}$ and $\gamma$ as they are experimentally measurable~\cite{Dagdigian_1971}. 
Figure~\ref{fig: FIG4} shows the variation of CBS values of average and anisotropic components of DP for all the molecules at the CCSD(T) level of theory. The average polarizability increases from AlF$^+$ to AlTs$^+$, and the anisotropic polarizability follows the same trend, except for AlCl$^+$. \\

\begin{figure}[]
    \includegraphics[width=9.3cm,height=6.5cm]{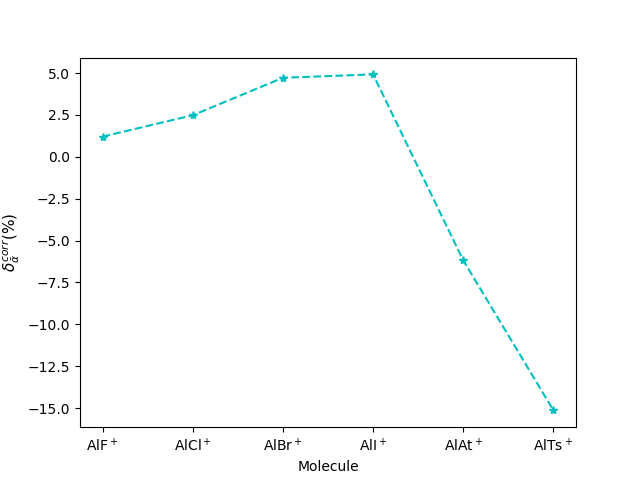}
    \caption{\label{fig:FIG5}Figure showing the relative percentage changes in the $\bar{\alpha}$ values of the \ce{AlX^+} (X: F, Cl, Br, I, At and Ts) molecular systems due to the electron correlation effects. }
\end{figure}

\begin{figure}[]
    \includegraphics[width=9.3cm,height=6.5cm]{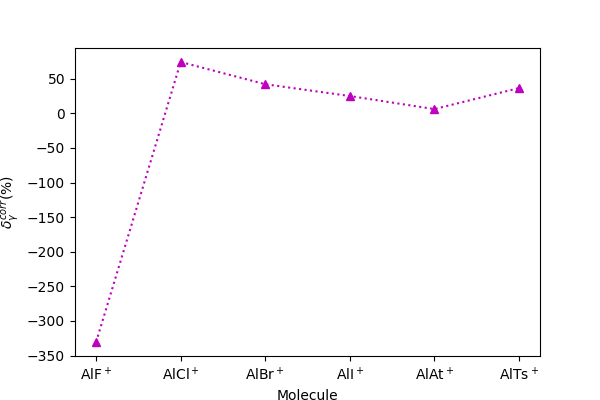}
    \caption{\label{fig:FIG6}Figure showing the relative percentage changes in the $\gamma$ values of the \ce{AlX^+} (X: F, Cl, Br, I, At and Ts) molecular systems due to the electron correlation effects. }
\end{figure}
We start by discussing the correlation effects, based on Figures \ref{fig:FIG5} and \ref{fig:FIG6}, which show the fractional percentage  difference in the values of $\bar{\alpha}$ (denoted as $\delta^{\text{corr}}_{\text{${\bar{\alpha}}$}}$) and $\gamma$ (denoted as $\delta^{\text{corr}}_{\text{${\gamma}$}}$), respectively, due to correlation effects. Electron correlation effects cause a decrease in the $\bar{\alpha}$ value for heavier molecular ions like AlAt$^+$ and AlTs$^+$, while resulting in an increase in the $\bar{\alpha}$ value for the remaining molecular ions. These effects lead to an increase in the $\gamma$ value for all molecular ions, with AlCl$^+$ showing the highest fractional percentage difference between CCSD(T) and DHF results, approximately \(74\)\%. Figure~\ref{fig:FIG6} shows a significantly different value of $\delta_{\gamma}^{corr}$ for AlF$^+$ molecular ion compared to the other molecular ions. This discrepancy is due to the large difference in the value of $\gamma$ at the DHF and CCSD(T) levels of theory. We have discussed the reason for the discrepancy in values in the later part of the manuscript. \\

\begin{figure}[ht]
    \includegraphics[width=9.3cm,height=6.5cm]{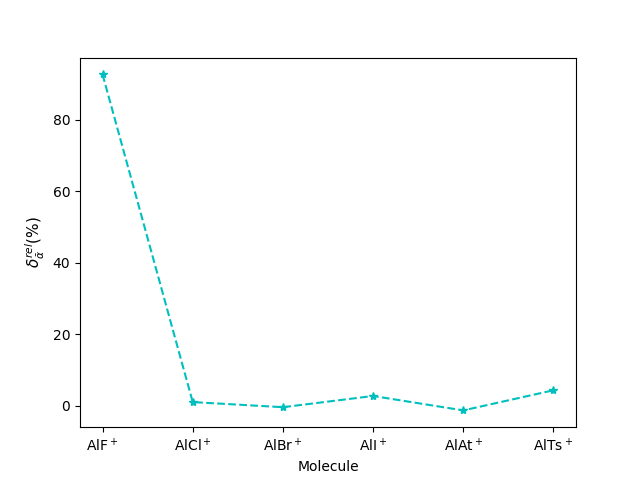}
    \caption{\label{fig:FIG7}Figure showing the relative percentage changes in the $\bar{\alpha}$ values of the \ce{AlX^+} (X: F, Cl, Br, I, At and Ts) molecular systems due to the electron relativistic effects. }
\end{figure}

\begin{figure}[ht]
    \includegraphics[width=9.3cm,height=6.5cm]{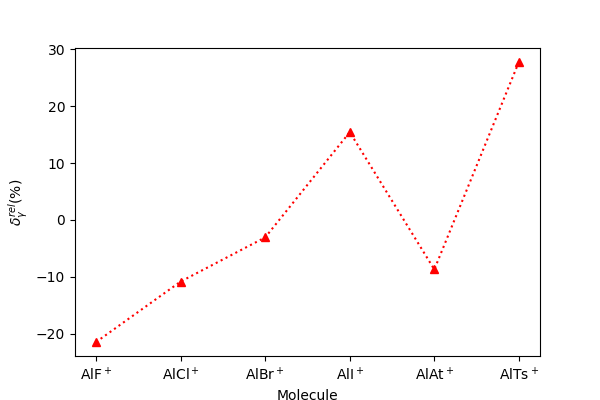}
    \caption{\label{fig:FIG8}Figure showing the relative percentage changes in the $\gamma$ values of the \ce{AlX^+} (X: F, Cl, Br, I, At and Ts) molecular systems due to the electron relativistic effects. }
\end{figure}
After discussing the correlation effects, we now address the effects of relativity on the components of static DP. We compare the relativistic and non-relativistic results of $\bar{\alpha}$ and $\gamma$ for all the molecules, as shown in Table~\ref{table-V}. Additionally, the fractional percentage  difference in the values of $\bar{\alpha}$ (denoted as $\delta^{\text{rel}}_{\text{$\bar{\alpha}$}}$) and $\gamma$ (denoted as $\delta^{\text{rel}}_{\text{${\gamma}$}}$) due to relativistic effects is plotted in Figures~\ref{fig:FIG7} and \ref{fig:FIG8}, respectively. The relativistic effects decrease the value of $\bar{\alpha}$ for AlBr$^+$ and AlAt$^+$, while increasing it for the rest of the molecular ions. These effects result in an increase in the $\gamma$ value for AlI$^+$ and AlTs$^+$, while reducing it for the other molecular ions, with the most significant increase, at \(28\)\%, observed for the AlTs$^+$ molecular ion. \\

\begin{table*}[t]
    \caption{\label{table-VI}Magnitude of permanent dipole moments (in Debye) and components of static dipole polarizability (in au) for \ce{AlX^+} (X: F, Cl, Br, I, At and Ts) molecular systems at different levels of correlation using singly augmented quadruple zeta (s-aug-dyall.v4z) basis sets.}
    \begin{ruledtabular}
    \begin{tabular}{c c c c c c c}
       Molecule & Method & $\mu$ & $\alpha_\parallel$ & $\alpha_\perp$ & $\bar{\alpha}$ & $\gamma$ \\
       \hline
       \ce{AlF^+} & DHF & 2.547 & 16.888 & 21.965 & 20.273 & -5.077 \\
                  & CCSD & 2.297 & 19.168 & 21.982 & 21.044 & -2.814 \\
                  & CCSD(T) & 2.219 & 19.795 & 22.100 & 21.332 & -2.305 \\
        \hline
        \ce{AlCl^+} & DHF & 0.740 & 38.611 & 31.287 & 33.728 & 7.324 \\
                    & CCSD & 0.349 & 43.075 & 31.415 & 35.302 & 11.660 \\
                    & CCSD(T) & 0.222 & 44.273 & 31.730 & 35.911 & 12.543 \\
        \hline
        \ce{AlBr^+} & DHF & 0.483 & 51.815 & 36.609 & 41.678 & 15.206 \\
                    & CCSD & 0.031 & 57.397 & 36.711 & 43.606 & 20.686 \\
                    & CCSD(T) & 0.185 & 58.931 & 37.125 & 44.394 & 21.806 \\
        \hline
        \ce{AlI^+} & DHF & 1.504 & 80.609 & 46.998 & 58.202 & 33.611 \\
                   &  CCSD & 2.333 & 88.634 & 47.197 & 61.009 & 41.437 \\
                   & CCSD(T) & 2.512 & 98.517 & 48.095 & 64.902 & 50.422 \\
        \hline
     \ce{AlAt^+}  & DHF & 4.307 & 109.140 & 61.718 & 77.525 & 47.422 \\
                  & CCSD & 4.293 & 103.710 & 60.028 & 74.589 & 43.682 \\
                  & CCSD(T) & 4.309 & 106.039 & 55.382 & 72.268 & 50.657 \\
        \hline
    \ce{AlTs^+} &  DHF & 6.729 & 136.359 & 93.799 & 107.986 & 42.560 \\
                &  CCSD & 6.575 & 131.980 & 62.110 & 85.400 & 69.870 \\
                &  CCSD(T) & 6.687 & 133.300 & 60.291 & 84.627 & 73.009 \\
    \end{tabular}
    \label{tab:my_label}
 \end{ruledtabular}   
\end{table*}
Similar to PDM, we have also checked the effect of diffuse functions on the components of static DP, utilizing singly augmented quadruple zeta basis sets. Table~\ref{table-VI} shows the outcomes for DPs using singly augmented basis sets. The augmentation of the basis sets does not notably alter the values of $\bar{\alpha}$ and $\gamma$ for molecular ions, except in the case of AlF$^+$, where there is a significant difference in the value of $\gamma$ at CCSD(T) level of theory. Table~\ref{table-IV} shows that the AlF$^+$ molecular ion exhibits a negative value of $\gamma$ at the DHF and CCSD levels of theory; however, the value becomes positive at the CCSD(T) level of theory. Table~\ref{table-VI} illustrates that the value of $\gamma$ adheres to the expected trend at different levels of correlation, emphasizing the importance of augmenting the basis sets for the AlF$^+$ molecular ion. The negative values of $\gamma$ have been reported in the literature for alkaline-earth monofluoride molecules in Ref.~\cite{Davis_1988,Bala_2019}.\\

We now discuss the possible sources of inaccuracies in our computations of static DP in the same way as that of PDM. We have found that increasing the virtual cut-off from \(12\) $E_h$ to \(25\) $E_h$ leads to a \(1.1\)\% change for $\bar{\alpha}$ and a \(0.4\)\% change for $\gamma$, while including \(49\) active electrons changes $\bar{\alpha}$ by \(1\)\% and $\gamma$ by \(1.9\)\%. Furthermore, the addition of augmented functions to the basis sets changes $\bar{\alpha}$ by \(0.3\)\% and $\gamma$ by \(0.6\)\%. The error resulting from the missing triple excitations in the CCSD(T) method would be about \(2.1\)\% for $\bar{\alpha}$ and \(13\)\% for $\gamma$. Hence, we anticipate an approximate \(4.5\)\% deviation for $\bar{\alpha}$ and a \(16\)\% deviation for $\gamma$ in our results.
\section{\label{sec:secIV}Conclusion}
In summary, we have done fully relativistic calculations for the PDMs and the components of static DPs of singly charged aluminum monohalide ions at the DHF and CCSD(T) level of theory using a hierarchy of progressively larger Dyall basis sets. Our findings for the molecular PDMs for the ground electronic state of aluminum monohalides agree well with the existing values in the literature. We have reported the results of molecular DPs for the first time in the literature to the best of our knowledge. The roles of relativistic and correlation effects in determining the PDMs and static DPs have been studied extensively in this work. We found that correlation effects impact the precision of final PDM values, with AlI$^+$ showing the maximum change at \(62\)\%, while the influence of relativity is particularly striking in superheavy AlTs$^+$, exhibiting the largest variation at \(50\)\%. We also observe that electron correlation tends to enhance $\bar{\alpha}$ and $\gamma$ for most molecules, whereas relativistic effects have varying impacts on these properties across different molecular ions. The effect of singly augmenting the basis sets on the molecular properties of the ground state is also investigated. Finally, we selected AlAt$^+$ as our representative system to evaluate errors, and our findings indicate that various sources of errors contribute together by about \(5.5\)\% for the PDM, \(4.5\)\% for $\bar{\alpha}$, and \(16\)\% for $\gamma$. We anticipate that our relativistic calculations, computed with large basis sets and large active space, would be beneficial for experimental spectroscopists interested in working with these molecular systems in the future.\\

\begin{acknowledgments}
We would like to thank the National Supercomputing Mission (NSM) for providing computing resources of ‘PARAM Ganga’ at the Indian Institute of Technology Roorkee, implemented by C-DAC and supported by the Ministry of Electronics and Information Technology (MeitY) and Department of Science and Technology (DST), Government of India. R.B. was supported by Polish National Science Centre Project No. 2021/41/B/ST2/00681. The research is also a part of the program of the National Laboratory FAMO in Toruń, Poland.
\end{acknowledgments}
\bibliography{AlX+_ions}

\newpage
\onecolumngrid
\appendix
\end{document}